\documentclass[10pt,letterpaper,twocolumn]{article}
\usepackage[latin9]{inputenc}
\usepackage{amsmath}
\usepackage{amssymb}
\usepackage{graphicx}
\usepackage[unicode=true,pdfusetitle,
 bookmarks=true,bookmarksnumbered=false,bookmarksopen=false,
 breaklinks=false,pdfborder={0 0 1},backref=section,colorlinks=false]
 {hyperref}
\hypersetup{
 draft}
\usepackage{breakurl}

\makeatletter

\usepackage{ol2}

\makeatother

\begin{document}
\twocolumn[
\title{Helico-conical optical beams self-heal}
\author{N. Hermosa,$^{1,*}$ C. Rosales-Guzm\'an,$^1$ and J.P. Torres$^{1,2}$}
\address{
$^1$ICFO - Institut de Ciencies Fotoniques, UPC, Mediterranean Technology Park \\ 08860 Castelldefels (Barcelona), Spain
\\
$^2$Department of Signal Theory and Communications, UPC, 08034
Barcelona, Spain

$^*$Corresponding author: nathaniel.hermosa@icfo.es
}

\begin{abstract} An optical beam is said to be self-healing when, distorted by an obstacle, the beam corrects itself upon
propagation. In this letter, we show through experiments supported
by numerical simulations, that Helico-conical optical beams
(HCOBs) self-heal. We observe the strong resilience of these beams
with different types of obstructions, and relate this to the
characteristics of their transverse energy flow.
\end{abstract}


 ]

%% activate for two-column option

\noindent %% activate for two-column option

\noindent The study of the self-healing properties of beams is of
great interest in optics\cite{Vyas2011,Paity2011,Broky2008,Morales2007,Ring2012,Bouchal2002}.
An optical beam is said to be self-healing when, after propagation,
its transverse intensity profile is hardly affected by a small perturbation
- a block - has been placed in its path\cite{Bouchal2002,Broky2008,Morales2007,Paity2011,Ring2012}.
The surge of interest in self-healing beams is bouyed mainly by its
range of applications; self-healing can be advantageous, for instance,
in beam propagation through scattering and turbulent media, and in
optical manipulation\cite{Garc=00003D00003D0000E9s-Ch=00003D00003D0000E1vez2002,Baumgartl2008}.

Optical beams that exhibit self-healing include Bessel beams (BBs)\cite{Durnin1987,Garc=00003D00003D0000E9s-Ch=00003D00003D0000E1vez2002,Vyas2011},
caustic beams\cite{Morales2007}, Airy beams \cite{Baumgartl2008,Broky2008},
Pearcey beams \cite{Ring2012}, the non-paraxial Mathieu and Weber
accelerating beams\cite{Zhang2012}, and some forms of Laguerre-Gaussian
(LG) beams\cite{Paity2011}. In the case of the BBs and Airy beams,
self-healing happens at a relatively small propagation distance, while
LG beams self-heal at a distance of the order of the Rayleigh length\cite{Paity2011,Ring2012}.
Self-healing is independent of the diffracting nature of the beams,
as shown by caustic \cite{Morales2007} and LG beams\cite{Paity2011}.

In this Letter, we present another set of beams that self-heal: the
Helico-conical optical beams (HCOBs). The main difference between
these beams and other self-healing beams is the non-separability of
their radial and azimuthal phases\cite{Alonzo2005}. HCOBs posses
a phase $\psi$ that is the product of a helical phase and a conical
phase: $\psi(r,\theta)=\ell\theta(K-r/r_{0}),$ where $\ell$ is the
winding number around the azimuth angle $\theta$, $r_{0}$ normalizes
the radial coordinate $r$, and $K$ takes either the value 0 or 1.
At the far-field, the intensity profile of these beams resembles a
spiral, with $K=1$ HCOBs having a more pronounced head near the center
of the beam axis compared with the $K=0$ HCOBs. Recently, a $K=0$
HCOB was reported to cause a spiral motion to a particle along its
path\cite{Daria2011}, a three dimensional motion that combines phase
gradient with intensity gradient forces\cite{Padgett2011}.

It could be argued that since HCOBs have conical phases, they should
behave similar to BBs. In fact, the HCOBs are more likely to be compared
with fractional higher-order BBs, because of their similar intensity
distributions\cite{Tao2004}. Joint to this is the fact that HCOBs
consist of strings of optical vortices upon propagation\cite{Hermosa2007}.

However, the far-field intensity pattern of experimentally generated
Bessel beams, or any superposition of it, resembles a circle or a
$\delta$-ring\cite{Thomson2007}, while HCOBs are spirals in the
far field, and not rings when compared to BBs \cite{Alonzo2005}.
An important question then arises: \emph{Can HCOBs self-heal?}

Here, we provide evidence that an HCOB reconstruct its intensity profile
at a relatively short propagation distance after a small perturbation
is placed in its path. We observe how the beam reconstructs for different
values of $\ell$ and for different block sizes. Since the phase of
the HCOB is not rotationally symmetric, we also note how the beam
reconstruct when we change the orientation of the obstructing block.
We then compare our experimental results with numerical simulations.
Finally, we look at the transverse energy flow of the beam and relate
it to its self-healing property.

\begin{figure}[tb]
\centerline{\includegraphics[width=7.5cm]{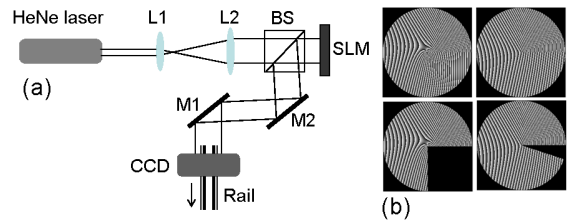}} \caption{(a) The experimental setup. Holograms are encoded onto a SLM. (b)
Samples of unblocked (above) and blocked (below) holograms are shown.
L1 and L2 are collimating lenses while M1 and M2 are mirrors for alignment.
BS is beam splitter.}
\end{figure}

We generate the HCOBs using a spatial light modulator (SLM)\cite{Vasara1989,Chattrapiban2003}.
Figure 1(a) shows the experimental set-up. A collimated HeNe laser
($\lambda=632.8nm$) beam impinges onto an SLM (Hamamatsu LCOS-SLM)
encoded with a computer-generated hologram. The holograms ($r_{0}=2.5$
mm) are calculated from the phase of the HCOBs. A carrier frequency
is added to separate the beam of interest. We observe the beam after
propagation from the SLM while varying $\ell$, and the size and orientation
of the block. We imitate the presence of a block by means of an incomplete
hologram (see Fig. 1(b)). This is done for better control of the size
of the block. The block size corresponds to an angular fraction $\Delta\theta$.
We then capture the intensity pattern with a charged couple device
(CCD) camera attached to a computer. Using the Angular Spectrum Method\cite{Goodman2005},
we compare our results with numerical simulations. Figure 2 shows
the intensity profiles of the unblocked and blocked HCOBs with $\ell=50$
and $\Delta\theta=\pi/3$. The intensity patterns, not measured in
the far-field, scale with the value of $\ell$, similarly to what
is shown in \cite{Alonzo2005}.

\begin{figure}[tb]
\centerline{\includegraphics[width=7cm]{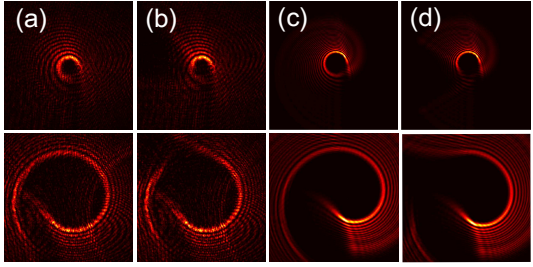}} \caption{Intensity profiles (false color, color online) of $\ell=50$ HCOBs
where (a) and (b) are obtained from experiments while (c) and (d)
are from numerical simulations. The top images are for $K=0$ while
the bottom ones are for $K=1$. The block size used in (b) and (d)
is $\Delta\theta=\pi/3$. }
\end{figure}

\begin{figure}[tb]
\centerline{ \includegraphics[width=6cm]{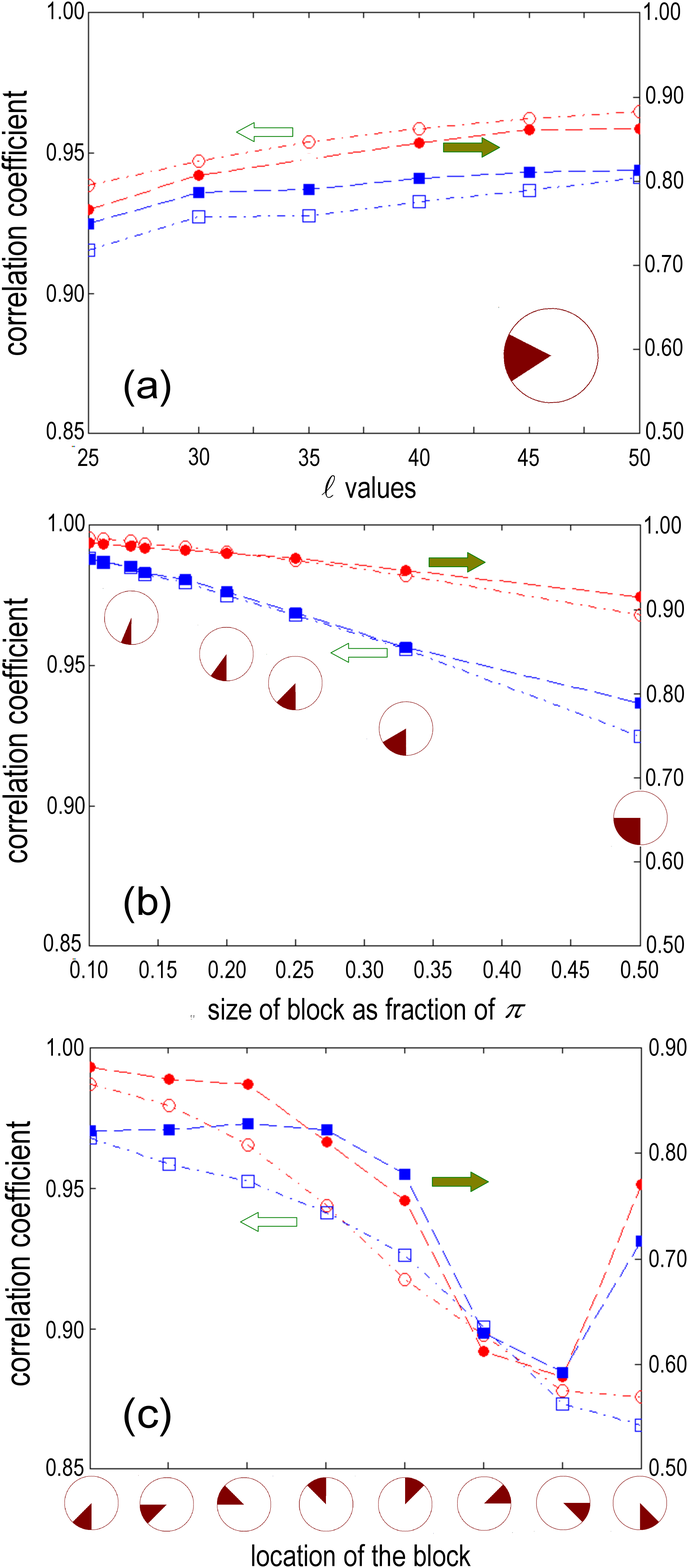}} \caption{(color online) (a) Correlation for different values of $\ell$. (b)
Correlation with different block sizes. (c) Correlation for different
block orientations. Numerical simulation (open points) uses the axis
on the left, while the right axis is for experiments (filled points).
Circle are for $K=0$, and square for $K=1$.}
\end{figure}

The similarities between blocked and unblocked beams are quantified
for both experiments and simulations using 2D image correlation \cite{Matsumoto2008}.
The value of the correlation coefficient ranges from $0$ for non-identical
beams to $1$ for identical beams . We emphasize, however, that the
correlation coefficient only gives a trend. It \emph{is not} an exact
measure of the quality of the beam reconstruction, especially in our
case, wherein it is difficult to separate the generated beam from
the adjacent diffraction orders.

The correlation coefficient changes with the value of $\ell$ as shown
in Fig. 3(a). The block size is $\pi/3$ and the orientation of the
block is shown as an inset. As the $\ell$ value increases, the correlation
coefficient also increases in both the experiment and numerical simulations,
with the $K=0$ HCOBs having higher correlation coefficients than
the $K=1$ HCOBs with the same $\ell$, at the same propagation distance.

The block size affects the reconstruction of the HCOBs as shown in
Fig. 3(b). The $K=0$ HCOBs reconstruct faster than the $K=1$ HCOBs,
given the same block size and the same propagation distance. We notice
that the beam with a $\pi$ block size (not shown) gives a very low
correlation coefficient which is consistent with the results previously
reported for other self-healing beams\cite{Morales2007,Vyas2011}.
The HCOBs reconstruction also depends on the orientation of the block
as shown in Fig. 3(c). The block size is $\Delta\theta=\pi/4$ and
the orientation of the block is shown below the plot. This is expected
since the intensity and the phase of the beam are not rotationally
symmetric.

\begin{figure}[tb]
\centerline{ \includegraphics[width=7cm]{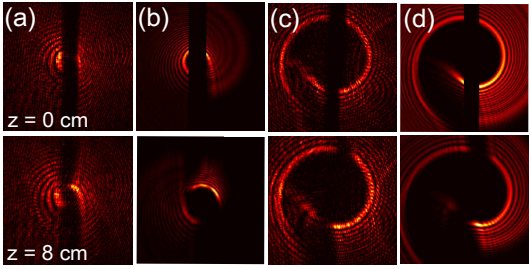}} \caption{(false color, color online) A $0.38mm$ strip is placed at the path
of a $\ell=40$ HCOB, $16cm$ after the SLM. (a) and (c) are experimental
results while (b) and (d) are numerical simulations.}
\end{figure}

In another experiment, we block the HCOBs with a $0.38$ mm opaque
strip, and observe how the HCOBs reconstruct upon propagation. We
restrict our measurement to distances below $r_{0}^{2}/\ell\lambda$,
since we notice that beyond this distance, the HCOBs' intensity profiles
change more rapidly\cite{comment1}. Figure 4 shows the experimental
and numerical results, with an opaque strip placed $16$ cm after
the SLM. Placing the CCD at different distances, we observe that the
HCOBs self-heal as the beams propagate. The shadow of the block moves
in a rotatory manner, similar to the self-healing of higher order
BBs. However, unlike BBs, HCOBS slightly rotate and expand.

\begin{figure}[tb]
\centerline{ \includegraphics[width=8cm]{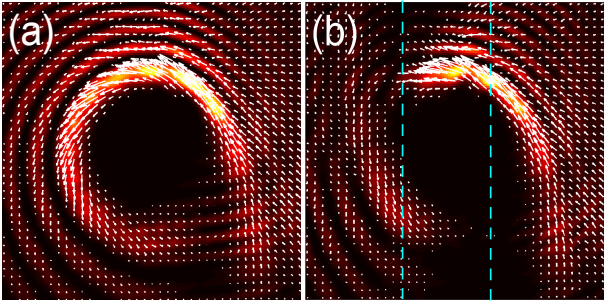}} \caption{(false color, color online) Transverse energy flow for $K=0$ HCOBs
with $\ell=30$ for (a) no block, and (b) blocked located $16cm$
after the SLM. Both beams propagate a total distance of $20cm$ .
Dashed lines denote the original position of the block.}
\end{figure}

Figure 5 shows the transverse energy flow of an $\ell=30$ $K=0$
HCOB, calculated numerically \cite{Allen1992}. Similar images can
be obtained for $K=1$ HCOBs. The direction of the energy flow traces
a curved path (shown as white arrows in the figure). Even with the
block, the direction of the energy flow is unaltered and so, the energy
flows from the surrounding areas to the blocked area\cite{Bekshaev}.
Since the energy and the energy flow are greater at the upper section,
the beam reconstructs faster in this part. In addition, the transverse
energy flow is greater for larger $\ell$ values which translates
to faster reconstruction. This suggests that the transverse energy
flow is the reason that the beam reconstructs.

In summary, we have shown experimentally that Helico-conical optical
beams self-heal and we have supported our results with numerical simulations.
The intensity profile reconstructs under different circumstances:
by varying the size and the orientation of the obstructing block,
as well as by changing the $\ell$ value of the HCOBs. We observe
how the beams heal as they propagate, and we link our results with
the transverse energy flow within the beam.

This work was supported by the Government of Spain (FIS2010-14831),
PHORBITECH (grant number: 255914) and Fundacio Privada Cellex Barcelona.
The authors would like to thank V. Rodriguez-Fajardo and L. J. Salazar-Serrano
for useful discussions.

%\pagebreak

\pagebreak{}

\section*{Informational Fourth Page}

In this section, please provide full versions of citations to assist
reviewers and editors (OL publishes a short form of citations) or
any other information that would aid the peer-review process.

\end{document}